\documentclass[twocolumn,aps,pra]{revtex4}
\usepackage{amsmath,amssymb}
\usepackage{graphicx}
\usepackage{epstopdf}
\usepackage[usenames]{color}

\newcommand{\be}{\begin{equation}}
\newcommand{\ee}{\end{equation}}
\newcommand{\bk}{{{\bf{k}}}}

\newcommand{\bQ}{{{\bf{Q}}}}
\newcommand{\br}{{{\bf{r}}}}

\newcommand{\hx}{{\hat{x}}}
\newcommand{\hy}{{\hat{y}}}

\newcommand{\bea}{\begin{eqnarray}}
\newcommand{\eea}{\end{eqnarray}}

\newcommand{\dg}{{\dagger}}
\newcommand{\pdg}{{\phantom\dagger}}

\begin{document}

\title{Use of Quantum Quenches to Probe the Equilibrium
Current Patterns of Ultracold Atoms in an Optical Lattice}

\author{Matthew Killi$^1$ and Arun Paramekanti$^{1,2}$}
\affiliation{$^1$Department of Physics, University of Toronto, Toronto, Ontario, Canada M5S 1A7}
\affiliation{$^2$Canadian Institute for Advanced Research, Toronto, Ontario, M5G 1Z8, Canada}
\begin{abstract}
Atomic bosons and fermions in an optical lattice can realize a variety
of interesting condensed matter states that support equilibrium current patterns in the presence
of synthetic magnetic fields or non-abelian gauge fields. As a
route to probing such mass currents, we propose
a nonequilibrium quantum quench of the Hamiltonian that dynamically converts the current patterns
into
experimentally measurable real-space density patterns. We illustrate how a specific
such ``unidirectional'' quench of the optical lattice can be used to uncover bulk checkerboard 
and stripe current orders in
lattice Bose superfluids
and Fermi gases, as well as chiral edge currents in a quantum Hall state.
\end{abstract}

\maketitle
Quantum dynamics of particles moving in magnetic fields or non-abelian gauge fields is of
great interest in various condensed matter systems such as the cuprate
superconductors \cite{leewen.rmp}, quantum Hall liquids \cite{stone.book}, and
topological insulators \cite{TI.review}.
Such gauge fields can result in equilibrium
charge or spin currents of electrons.
For instance,
a type-II superconductor in
a magnetic field forms an Abrikosov vortex lattice which supports a periodic 
bulk current pattern \cite{abrikosov}.
A uniform magnetic field for lattice electrons
can lead to
topologically nontrivial states with a quantized
Hall conductance and chiral edge currents \cite{tknn}. In a solid,
such electronic currents produce their own characteristic magnetic fields, and
can thus be probed by using magnetic
microscopy \cite{magmicroscopy} or neutron scattering \cite{greven}.

Recently, experiments have begun exploring effects of  ``artificial'' 
orbital magnetic fields \cite{spielman.nature2009,bloch.prl2011,dalibard.rmp,sengstock.prl2012}
and non-abelian gauge fields \cite{yjlin.nature2011} on neutral
ultracold atomic gases. These experiments can
potentially realize a wide variety of fermionic and bosonic states with equilibrium 
mass currents. This brings us to a crucial question of great interest: {\it
How can we observe the equilibrium mass current patterns for neutral 
atomic gases?}

In this paper, we propose a dynamical probe of currents, which
relies on experimental progress in measuring lattice scale modulations of the atom
density. Such density mapping tools include 
noise correlations \cite{noisetheory,noiseexpt}, Bragg scattering \cite{kuhr.prl2011}, and {\it in
situ} microscopy at the lattice scale \cite{greiner.nature2009}. 
Our key idea is to make a specific quantum quench of the Hamiltonian 
that violates the steady state {\it divergence-free} condition on equilibrium 
currents,  thus causing an imbalance between currents entering and leaving different
sites. This leads to characteristic density 
build-up or depletion, as dictated by the continuity equation. 
Imaging the subsequent density variation across the lattice then 
yields {\it real space}
information about the initial currents. 
Our proposal thus links up with another very important thread in ultracold atom research:
quench-induced quantum dynamics in many-body systems \cite{nondyn}.

Consider a thought experiment in which we start in an equilibrium state with
currents and 
instantaneously turn off the
particle hopping amplitude on every bond except one. Then, the instantaneous current ${\cal J}$ 
on that
one bond is unaffected, but currents on all other bonds vanish. A short time $\delta t$ after this
local quench, there will be an accumulated density imbalance ${\cal J} \delta t$ between the two
sites connected by the unperturbed bond; measuring this yields information on the magnitude and
direction of the bond current. A similar such ``quasi-local'' current probe has been used in a recent
experimental study of nonequilibrium dynamics in a 1D Bose gas \cite{trotzky}.

In this paper, we study equilibrium current patterns in 2D, focusing on  a specific
``unidirectional'' quench, where we suddenly decrease the 
particle hopping amplitude
in one direction across the entire lattice. This is achieved simply by changing the optical lattice
laser intensity along one direction. We show that tracking the subsequent space-time variation of
the density for this quench protocol can uncover the underlying
current patterns in many interesting cases which have been chosen to show that
the method works for both bosons and fermions, and can probe bulk as well as edge currents.
Specifically, we illustrate this scheme for
(i) a square lattice Bose superfluid in a staggered (checkerboard) magnetic flux
pattern, (ii) a square lattice Bose superfluid with a striped magnetic flux pattern, (iii)
noninteracting fermions on a square lattice in a staggered magnetic flux, and (iv) 
the integer quantum Hall state of lattice fermions in a
uniform magnetic flux that supports chiral edge currents. 
Such quenches can also probe the current pattern
in the recently studied chiral Bose Mott insulator \cite{dhar.pra2012} and spin currents of atomic
matter.

{\it Bosons in a checkerboard flux pattern. ---} 
Consider interacting bosons on a 2D square lattice,
described by the Bose Hubbard model $H_{\rm BH}=- \sum_{\br,\br'} J^\pdg_{\br,\br'} b^\dg_\br 
b^\pdg_{\br'}
+ \sum_\br V^\pdg_\br b^\dg_\br b^\pdg_\br + \frac{U}{2} \sum_\br b^\dg_\br b^\dg_\br b^\pdg_\br
b^\pdg_\br$. Here $J_{\br,\br'}$ denotes the boson hopping amplitude between sites $(\br,\br')$, $U$
is the on-site repulsion, and $V_\br$ is a harmonic trap potential given by $V_\br = V_0 (x^2+y^2)$.
We take $J_{\br,\br'} \! = \! J^*_{\br',\br} \! \neq \! 0$ only for nearest neighbors, and choose
$J_{\br,\br+\hat{x}} \! = \! J$ and $J_{\br,\br + \hat{y}} \! = \! J_y \exp (i (-1)^{x+y}
\phi/2)$. This yields staggered magnetic fluxes, $\pm \phi$, that pierce the elementary square
plaquettes in a checkerboard pattern; a route to realizing such a flux pattern has been
proposed previously \cite{hemmerich.stagsf}.

\begin{figure}[t]
	\includegraphics[width=0.48 \textwidth]{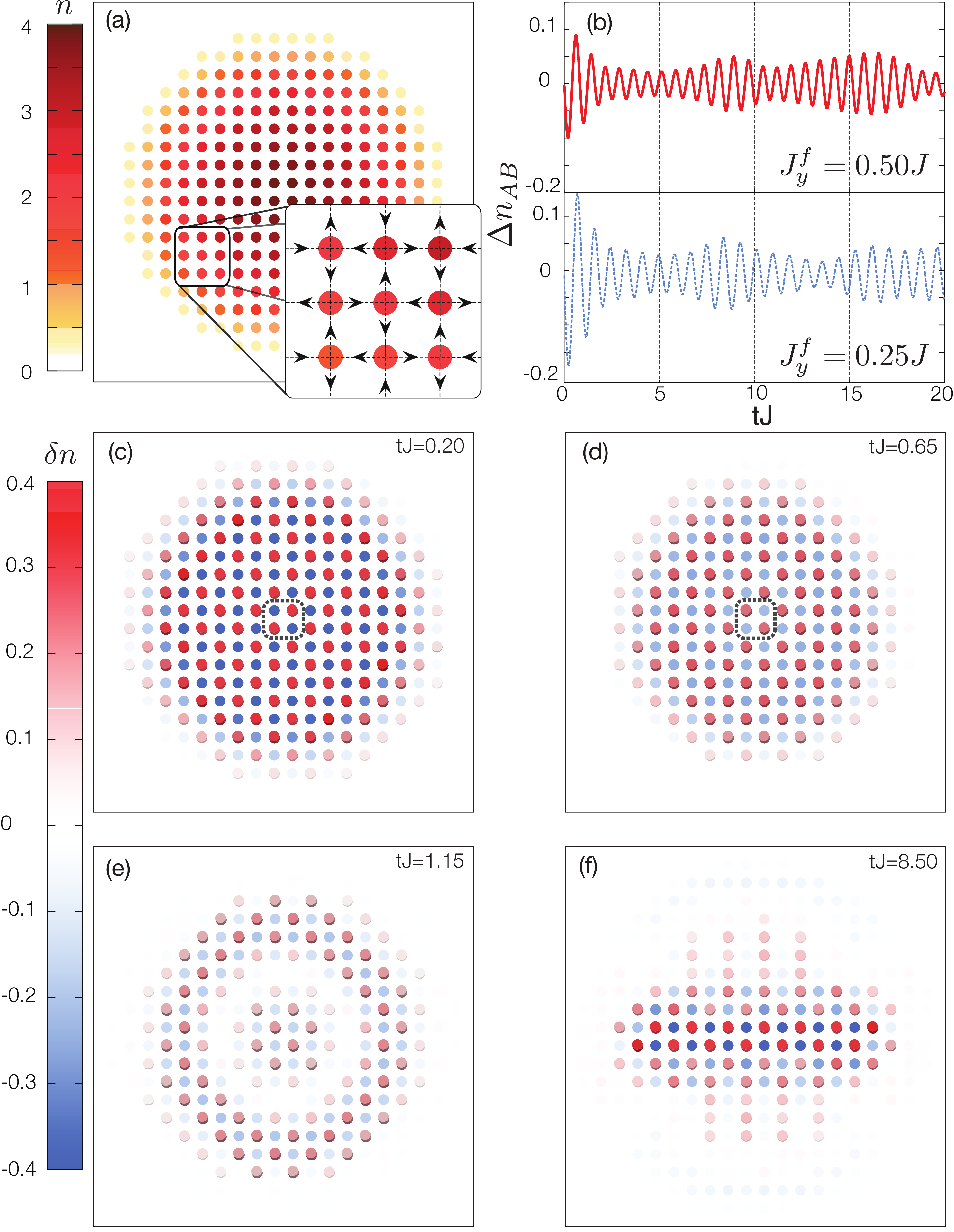}
	\caption {(Color online) Density pattern of 2D Bose superfluid in a staggered flux gauge field following a quench.  (a) Initial density profile and (inset) current pattern of condensate ground state at half-filling.  (b) Time dependence of the sublattice density difference $\Delta n (t)$ for a staggered flux of $|\phi|=\pi/2$ flowing quench of $J^i_y=J$ to $J^f_y=0.25J$ (dashed blue) and $J^f_y=0.50J$ (solid red).  (c -- f) Change in local density, $\delta n$ (relative to original density), at different times following a quench $J^i_y=J \to J^f_y=0.50J$ with 
$U=J$. Circled region in (c), (d) indicates central plaquette vortex.}
\label{Fig:boson_stag}
\end{figure}

For weak interaction, $U \! \lesssim \! J,J_y$, we first solve for the equilibrium ground
state \cite{supp} for $J_y\!\! =\! \! J$ by numerically minimizing the Gross-Pitaevskii (GP)
energy functional \cite{GPE} $E_{\rm GP} \! =\! - \sum_{\br,\br'} J^\pdg_{\br,\br'} \Psi^*_\br
\Psi^\pdg_{\br'} \! + \! \sum_\br (\frac{U}{2} |\Psi_{\br}|^4 + V_\br |\Psi_\br|^2)$, where $\Psi_\br$ 
is the condensate wavefunction at lattice site $\br \equiv (x,y)$. As shown in 
Fig.\ref{Fig:boson_stag}(a)
for a system with linear length $L=22$, $V_0=0.05J$, and an average filling factor of 
$\bar{n}=1$, this
leads to a superfluid ground state with staggered loop currents. The smooth density 
profile of the ground
state reflects the trap potential, but it does {\it not} reveal the currents induced by the gauge
field.

Starting with this ground state, we suddenly decrease $J_y$ from its initial value
$J_y^i \! = \! J$ to a final value $J_y^f \! < \! J$ at time $t\! =\! 0$. The subsequent
condensate dynamics is described by the time-dependent GP equation \cite{GPE}
\be 
i \hbar \frac{\partial
\Psi_\br(t)}{\partial t} = - \sum_{\br'} J_{\br,\br'}^f \Psi_{\br'}(t) + (U |\Psi_{\br}(t)|^2 \! +\!
V_\br) \Psi_\br(t).
\ee 
Henceforth, we set $\hbar=1$.
To study the dynamics, we numerically integrate this time-dependent
GP equation \cite{supp}, which allows us to extract the condensate
wavefunction $\Psi_{\br}(t)$, and the boson density $|\Psi_{\br}(t)|^2$, 
at later times.
As shown in
Fig.\ref{Fig:boson_stag}, the condensate at $t \! > \! 0$ develops a striking checkerboard {\it
density} pattern that reflects the underlying current order. Information about the direction
of circulation on a plaquette is easily discerned from the density pattern established after a short
time period; since the quench is along $J_y$, the initial build up of density is on sites that have
currents flowing into them along the strong $J$-bonds oriented along the $x$-axis. After a 
short time has passed, the density buildup reaches a maximum and the 
flow is
reversed, resulting in ``plasma oscillations'' between the two checkerboard patterns. The
frequency of these oscillations scales
as $\sim \sqrt{U J |\Psi_\br|^2}$; it varies slowly with position 
due to the density inhomogeneity in the trap.
Such checkerboard oscillations persist for a
long time, as inferred from the density difference between the two sublattices
$\Delta n_{AB}(t) = n_A(t) - n_B(t)$ shown in Fig.\ref{Fig:boson_stag}.

We find that the sublattice density pattern oscillations are also robust against
moderate random phase fluctuations imprinted on the initial state \cite{supp}.
This suggests that thermal phase fluctuations will suppress but not 
completely destroy these oscillations.
Remarkably, the sublattice density oscillation persists out to long times, $t J \gg 1$, as
seen in Fig.\ref{Fig:boson_stag}(b). At these times, shown in Fig.~\ref{Fig:boson_stag}(e) and
(f), we find additional
long-wavelength modulations superimposed on the checkerboard
density pattern. 
(i) Spherical
density waves emanate periodically outward from the centre, which we attribute to the spatial
variation of the ``plasma frequency'' resulting from the radial variation of
the
density $|\Psi_\br|^2$ in the trap. 
(ii) The
cloud shape shows oscillatory distortions into an ellipse due to the anisotropy of the
final tunneling $J_y < J$. 

\begin{figure}[t]
	\includegraphics[width=0.48 \textwidth]{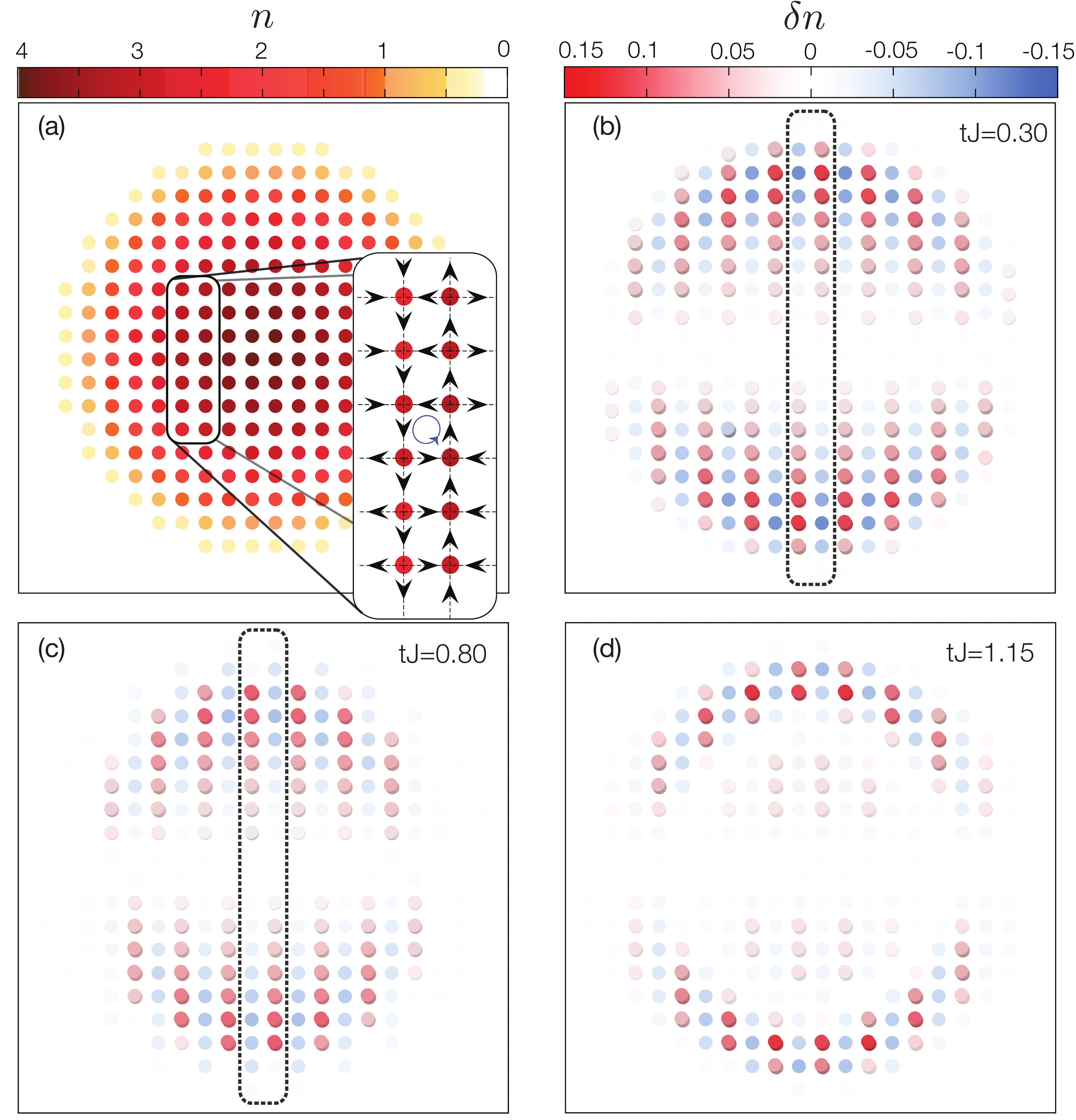}
	\caption {(Color online) Dynamical density pattern of interacting bosons on a 2D square lattice in a stripe-like magnetic flux pattern following a quench. (a): Initial density profile and (inset) current pattern of condensate ground state at half-filling.  (b -- e): Change in local density, 
	$\delta n$ (relative
	to the original density), at different times following a quench $J^i_y\!=\! J \to J^f_y\! 
	=\! 0.50J$ with $U\! =\! J$. Circled region indicates the central elongated vortex with a
	nonzero quadrupole moment.}
\label{Figboson_stripe}
\end{figure}

{\it Bosons in a stripe flux pattern. ---} We next
consider the above Bose Hubbard model in the presence 
of a striped magnetic flux pattern as realized in a recent experiment \cite{bloch.prl2011}.
We choose
$J_{\br,\br+\hat{x}} \! = \! J$ and $J_{\br,\br + \hat{y}} \! = \! J_y \exp (i
(-1)^{y} \phi x)$, so that we enclose fluxes $\pm \phi$ through each plaquette that lies 
along a stripe in the
vertical direction, and
solve for the equilibrium ground state by minimizing the GP
energy functional for $U \! \lesssim \! J,J_y$, with $L=22$, $V_0=0.05J$, and 
an average filling $\bar{n}=1$. We find a superfluid with vertically striped 
loop currents depicted in
Fig.\ref{Figboson_stripe}, 
which resembles a stripe pattern of `long vortices' that are 
highly elongated
along the $y$-direction. Again, the smooth equilibrium density pattern reveals {\it no} information about the underlying current order.

Upon quenching $J_y$, the superfluid generates a {\it density} pattern that
strikingly reflects the underlying equilibrium striped currents. Each vertically elongated loop
forms four quadrants of alternating high and low density, giving rise to an oscillatory quadrupole
moment  $\sum^{\rm loop}_{xy} x y \rho(x,y,t)$ about the center of each `long
vortex', where the sum is restricted to a single vertically elongated loop composed of square
plaquettes. The density pattern oscillates along the top (or bottom) half of the observed
density profile as we go along the $x$-direction. 
At intermediate times, we again find that the spatial dependence of the
oscillation frequency leads to additional long wavelength stripe-like waves 
emanating from the trap centre.
Quenching $J$ (i.e., along the $x$-direction) rather than $J_y$ leads to a similar early time 
density patterns. However the oscillatory dynamics occurs on longer time 
scales due to  mass transport occurring over larger distances $\sim L$.

{\it Spinless fermions in a staggered flux. ---} Motivated by exploring such quench-induced
density
dynamics for fermions, we next turn to noninteracting fermions in a 
staggered flux background \cite{stagflux1}. We study the Hamiltonian
$H_{\rm sf}  = - \sum_{\br,\br'} J^\pdg_{\br,\br'} 
f^\dg_\br f^\pdg_{\br'}$, where $J_{\br,\br+\hat{x}} \! = \! J$ and 
$J_{\br,\br + \hat{y}} \! = \! J_y \exp (i (-1)^{x+y} \phi/2)$, leading
to staggered checkerboard fluxes $\pm \phi$ \cite{hemmerich.stagsf}. 
In momentum space \cite{footnote.sf_freefermion},
$
H_{\rm sf}  = \sum_{\bk}^{\prime} \Omega_\bk
\Psi^\dg_\bk
(\cos\theta_\bk \tau^z + \sin\theta_\bk \tau^y)
\Psi^\pdg_\bk$,
where
$\Psi^\dg_\bk = (f^\dg_\bk, f^\dg_{\bk+\bQ})$, 
$\tau^{y,z}$ are Pauli matrices,
and the prime on the sum implies that only momenta
in the reduced Brillouin zone (BZ) are included.
Here, we have defined
$\Omega_\bk \! = \! \sqrt{\varepsilon^2_\bk + \gamma^2_\bk}$, $\cos\theta_\bk \! = \! 
\varepsilon_\bk/\Omega_\bk$,
and $\sin\theta_\bk \! = \! \gamma_\bk/\Omega_\bk$, with
$\varepsilon_\bk \! = \! - 2 (J \cos k_x + J_y \cos \frac{\phi}{2} \cos k_y)$ and
$\gamma_\bk \! = \!  - 2 J_y \sin \frac{\phi}{2} \cos k_y$. This
leads to mode energies $\pm \Omega_\bk$.

\begin{figure}[t]
\includegraphics[width=3in,height=2in]{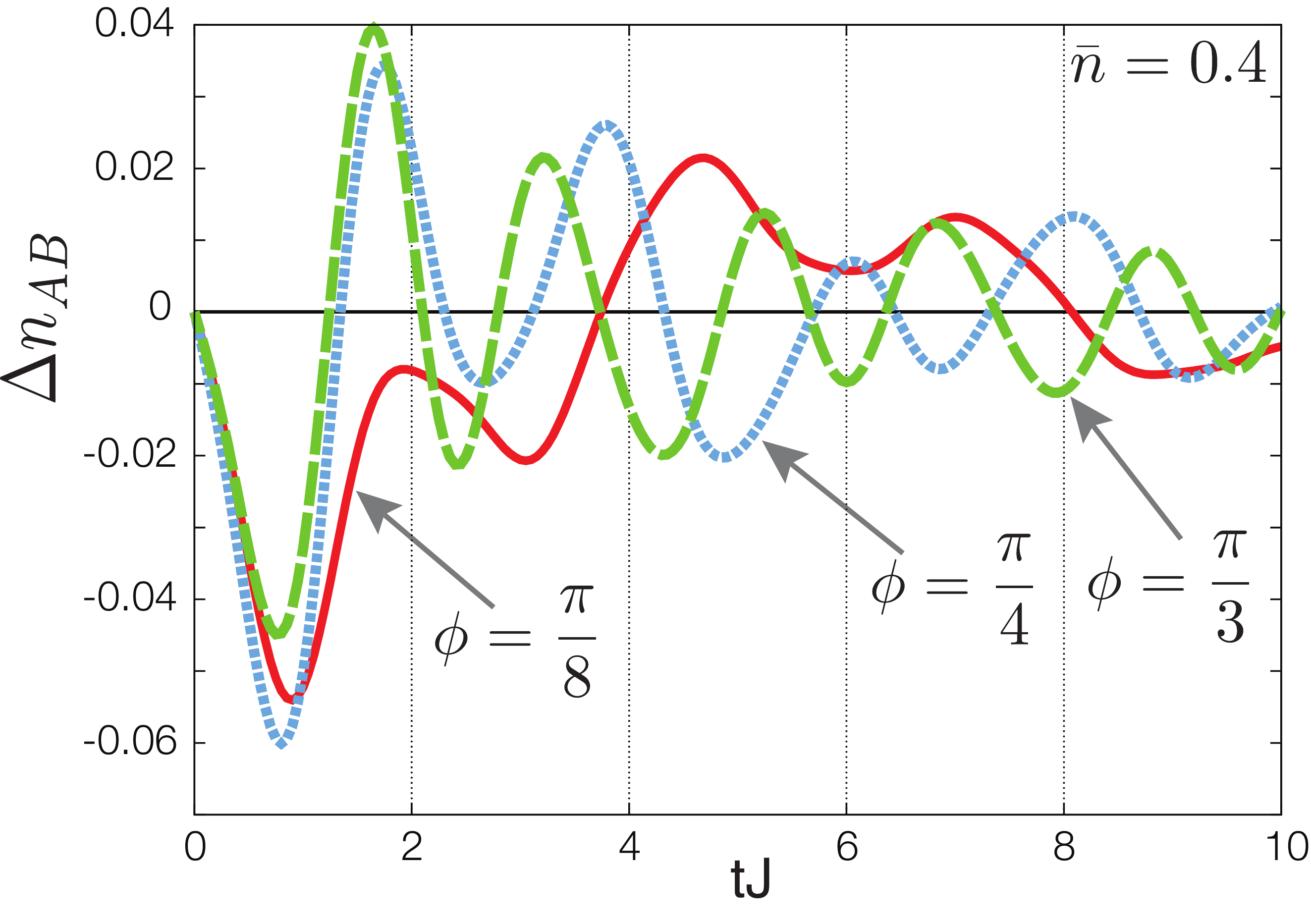}
\caption{(Color online) Time dependence of the sublattice density difference 
$\Delta n_{AB} (t)$, for noninteracting spinless fermions on a 2D square lattice at a
filling of $\bar n = 0.4$, and various staggered fluxes $\phi$,
following a
quench $J^i_y=J \to J^f_y=0.5J$.}
\label{Fig:fermion_free}
\end{figure}

Imagine fermions initially filled into negative energy states $-\Omega_\bk$ 
up to a Fermi energy $E_F$, and then
quenching $J_y$ from $J_y^i \! \to \! J_y^f$
at time $t\! =\! 0$, which instantaneously changes 
$(\Omega_\bk,\theta_\bk) \to (\tilde\Omega_\bk,\tilde\theta_\bk)$. 
This translationally invariant quench ensures that
different momentum pairs $(\bk,\bk+\bQ)$ stay 
decoupled from each other.
The
ensuing ``spin precession'' type dynamics in momentum space
leads to an oscillating
density difference between the two sublattices \cite{supp}
$\Delta n_{AB} (t) = \frac{1}{N} \sum_{\{\bk\}_{\rm occ}}^{~~~~~~\prime} \sin(\theta_\bk - \tilde\theta_\bk) 
\sin(2 \tilde\Omega_\bk t)$
where the momentum sum runs over only {\it initially occupied} states in the reduced BZ.
A numerical evaluation of the sum
allows us to plot the sublattice density oscillations, shown in 
Fig.~\ref{Fig:fermion_free} for $J_y^i \! = \! J$,
$J_y^f \! =\! 0.5 J$, fermion density $\bar n \! = \! 0.4$ per site, and various staggered
flux values. 

These oscillations exhibit multiple frequencies 
due to the large number of occupied fermion modes; the very 
generation of such a sublattice imbalance is evidence of an initial staggered current pattern.
Over the entire range of
displayed fluxes, and a wide range of
densities $\bar n \! \sim \! 0.3$-$0.5$ near half-filling, we find that
the dominant oscillation frequency arises from
initially occupied states
near $\bk\! =\! (0,\pi)$ due to a van Hove singularity in the density of states \cite{supp}.
This leads to an estimated, nearly density-independent, dominant
oscillation frequency
$\tilde\Omega^* \approx  2 \sqrt{J^2+(J^f_y)^2 - 2 J J^f_y \cos\frac{\phi}{2}}$,
in good agreement with the numerical data in Fig.~\ref{Fig:fermion_free}. 
The
weak density dependence of $\Delta n_{AB}(t)$ over a range of fillings indicates that trap 
induced inhomogeneities 
will not significantly affect these oscillations.

{\it Lattice quantum Hall state. ---} Finally, we turn to
topologically nontrivial states that
support {\it edge} currents. Recent work has focused on extracting
the nontrivial band topology from time-of-flight measurements \cite{zhaospielman,cooper} 
or spectroscopy of the edge modes \cite{gerbier}.
Here we explore
density dynamics induced by the unidirectional quench for lattice fermions in a uniform
magnetic field.
Proposals
to obtain such uniform fluxes exist in the literature \cite{cooper,trotzky}.
For concreteness,
consider fermions on a 2D square lattice
with a uniform magnetic flux $\phi\! \! = \! \! 2\pi/3$ per plaquette. 
The resulting particle-hole symmetric Hofstadter
spectrum \cite{hofstadter} has three non-overlapping bands, with
Chern numbers
$+1,-2,+1$, so that `band
insulators' with {\it some} bands being completely filled support a nonzero quantized 
Hall conductance, and chiral 
edge currents, yielding lattice quantum Hall (QH) states \cite{tknn}.

\begin{figure}
\includegraphics[width=3.4in]{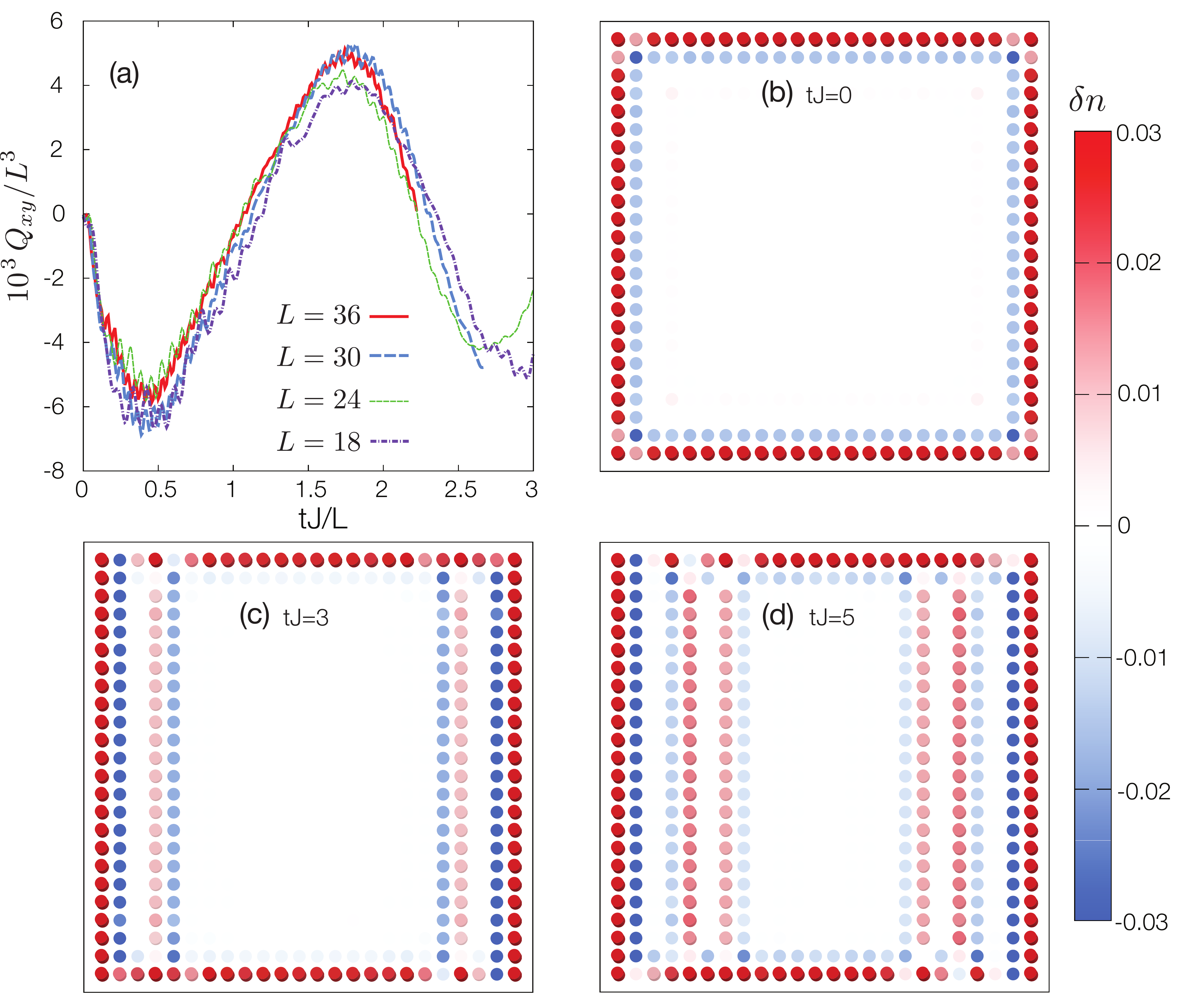}
\caption{(Color online) Density dynamics for
spinless fermions in the lowest Hofstadter band with flux $\phi=2\pi/3$
per plaquette on a square lattice following a quench
from $J_y^i=J$ to $J_y^f = 0.5 J$. (a) The scaled 
quadrupole moment 
$Q_{xy}(t)/L^3=(1/L^3) \sum_{x,y} x y \delta n(x,y,t)$
for various system sizes versus the scaled time $t J/L$.
(b -- d) Stripe-like density modulations (for $L=24$)
moving from the $y$-edges into the initially incompressible bulk
at different indicated times.}
\label{Fig:QHE}
\end{figure}

We begin by numerically diagonalizing the Hamiltonian 
$H_{\rm QH}  = - \sum_{\br,\br'} J^\pdg_{\br,\br'} 
f^\dg_\br f^\pdg_{\br'}$ with $J_{\br,\br+\hx} = J$ and $J_{\br,\br+\hy} = 
J_y {\rm e}^{i \phi x}$, 
for $\phi=2\pi/3$,
with open boundary conditions on a $L \times L$ system,
and fill up the lowest band (and some edge modes) to get a
fermion filling $\bar n=1/3$.
We find that the ground state bulk
density is uniform (see Fig.~\ref{Fig:QHE} for $t=0$) and supports edge currents 
confined to an ``edge layer'' of thickness $\sim 2-3$ lattice sites.
We next track the density 
dynamics following a quench from $J_y^i = J$ to  $J_y^f < J$, which is
easy to study once we compute the
initial and final spectrum and eigenstates.

Viewing the chiral edge currents as analogous to that arising from a `vortex',
we expect the
quench to lead to quadrupolar density 
oscillations and current reversals, similar to what we found for the
`long vortex' in the stripe flux superfluid. 
Inspired by recent work on the
superfluid Hall effect of atomic bosons \cite{Leblanc},
we study the behavior of the 
quadrupole moment $Q_{xy}(t)= \sum_{x,y} x y \delta n (x,y,t)$, 
where the density deviation $\delta n$ is with respect to
$\bar n\! =\! 1/3$. We find that $Q_{xy}(t)$ indeed displays oscillatory
sign reversals and, as seen in Fig.\ref{Fig:QHE}(a), the data for various $L$ 
collapse when
plotted as $Q_{xy}(t)/L^3$ versus $t/L$. The $t/L$ scaling shows that the 
oscillations occur due to transport across the system length $L$.
A simple scaling argument for an edge current induced oscillation shows
that $Q_{xy} \sim L^3$, as we also see numerically.
The numerical observations
are thus consistent with the quadrupolar oscillations being driven by edge currents.
For $\phi=-2\pi/3$, $Q_{xy}(t)$
has the opposite sign. Taken together, these observations provide strong
evidence that the initial state is a nontrivial insulator which is 
{\it incompressible in the bulk and supports 
chiral edge currents}.

We observe two additional effects. (i) There are
rapid small amplitude oscillations of $Q_{xy}(t)$, over a time scale $\sim J^{-1}$, 
superimposed on the long time sign reversals. (ii) 
The quench 
leads to the dynamic stripe-like density modulations
(see Fig.~\ref{Fig:QHE}) which
nucleate near the $y$-edges and propagate inward into the {\it bulk} from 
both directions.
We attribute both observations to quench-induced `band mixing'.
Such mixing occurs across the lattice, but the asymmetry at the edge 
leads to the stripes originating from the edge. Such
mixing with bands of opposite Chern number is also crucial for
edge current reversals \cite{supp}.

{\it Discussion. ---} 
We have shown that quantum quenches can yield a probe of 
underlying current patterns of atoms in an optical
lattice by 
converting them into measurable real-space density orders.
A possible
concern regarding such sudden quenches is that it may
invalidate our tight-binding 
model since particles can get excited to very high bands of the periodic optical
potential.
To address this concern,  the quench needs to be
``adiabatic'' on time-scales comparable to the inverse interband gap while being ``sudden''
on time scales governing intraband dynamics.
This seems to be achievable; for instance, time dependent lattice modulation experiments on 
repulsive atoms in deep lattices can access regimes where they are consistent with 
theory on the one-band Hubbard model \cite{doublon}.
Looking ahead, it would be extremely useful to explore
quench protocols for distinguishing various topological
states that may be realized in future experiments.

We thank L. LeBlanc, J. Thywissen, S. Trotzky, and E. Zhao 
for illuminating discussions,
and acknowledge funding from NSERC of Canada.


\begin{thebibliography}{51}
\bibitem{footnote.sf_freefermion}
To make analytical progress, we ignore the trap potential 
in this discussion.
\end{thebibliography}
\end{document}